\documentclass[aip,jcp,preprint
]{revtex4-1}
\usepackage{amsmath}    % need for subequations
\usepackage{graphicx}   % need for figures
\usepackage{xcolor}      % use if color is used in text
\usepackage{subfig}     % Teilbilder erzeugen
\usepackage{units,nicefrac}
\usepackage{hyperref}

\begin{document}
\title{Momentum dependence of the excitons in pentacene}
\author{Friedrich Roth}
\affiliation{IFW Dresden, P.O. Box 270116, D-01171 Dresden, Germany}
\author{Roman Schuster}
\affiliation{IFW Dresden, P.O. Box 270116, D-01171 Dresden, Germany}
\author{Andreas K\"onig}
\affiliation{IFW Dresden, P.O. Box 270116, D-01171 Dresden, Germany}
\author{Martin Knupfer}
\affiliation{IFW Dresden, P.O. Box 270116, D-01171 Dresden, Germany}
\author{Helmuth Berger}
\affiliation{Institute of Physics of Complex Matter, Ecole Polytechnique Federale de Lausanne (EPFL), CH-1015 Lausanne, Switzerland}

\date{\today}

\begin{abstract}
We have carried out electron energy-loss investigations of the lowest singlet excitons in pentacene at 20 K. Our studies allow to determine the
full exciton band structure in the $\bf{a^*},\bf{b^*}$ reciprocal lattice plane. The lowest singlet exciton can move coherently within this
plane, and the resulting exciton dispersion is highly anisotropic. The analysis of the energetically following (satellite) features indicates a
strong admixture of charge transfer excitations to the exciton wave function.
\end{abstract}

\maketitle

%%%%%%%%%%%%%%%%%%%%%%%%%%%%%%%%%%%%%%%%%%%%%%%%%%%%%%%%%%%%%%%%%%%%%%%%%%%%%%%%%%%%%%%%%%%%%%%%%%%%%%%%%%%%%%%%%%%%%%
\section{Introduction}

Pentacene as well as other members of the oligoacene family are famous organic semiconductors. Single crystals of these materials can be grown
in a very high quality, and thus very high charge carrier mobilities could be achieved in e.g. pentacene and tetracene \cite{Karl2003}. These
have been exploited in organic field effect transistors in view of fundamental as well as applied aspects
\cite{Gershenson2006,Takahashi2007,Braga2009,Sirringhaus2009}. Moreover, organic semiconductors are also of interest for manufacturing large
area and flexible devices for light emission or light harvesting \cite{Hepp2003,Santago2005,Friend2012,Schlenker2011,Meerheim2009}. In this
context, the electronic excitations in pentacene play an important role.

\par

The optical properties of pentacene and related oligoacenes have been studied previously with the aim to microscopically understand their
behavior and gain fundamental insight into the materials itself but also into the entire class of organic semiconductors. One subject of
particular interest is the dynamics of excitons in these structures as they are decisive for light absorption and emission properties
\cite{Pope}. A number of recent studies indicate that the fundamental singlet excitations in oligoacene and other $\pi$-conjugated molecular
crystals are of complex nature and cannot be rationalized on the basis of molecular Frenkel excitons only. Instead, a mixture of Frenkel and
charge transfer (CT) excitations has been discussed to be responsible for the observed behavior
\cite{Schlosser1980,Bounds1980,Sebastian1981,Petelenz1988,Slawik1999,Hoffmann2000,Petelenz1996,Knupfer2002,Hoffmann2002,Tiago2003,Knupfer2004,Lim2004,Petelenz2004,Hummer2005,Schuster2007,Gisslen2009,Gangilenka2010,Yamagata2011,Gisslen2011,Mazur2012,Stradomska2012}.
In particular, recent theoretical work suggests that neither the Davydov splitting, arising from the interaction of the two inequivalent
molecules in the unit cell of oligoacenes, nor the exciton dispersion can be understood without the inclusion of charge transfer excitons
\cite{Yamagata2011}. Further, also the spectral shape of the electronic excitation spectrum can depend substantially on the admixture of charge
transfer excitations \cite{Gisslen2009}. Moreover, for oligoacenes it has been predicted that the charge transfer contribution to the lowest
lying singlet excitons becomes more prominent for larger molecules \cite{Yamagata2011}. In view of this prediction and due to the fact that the
exciton dispersion in organic semiconductors determines the material's behavior in light emitting or absorbing devices, it is highly desirable
to know the exciton dispersion in such crystals. For instance, luminescence from exciton states at finite momentum in
N,N'-dimethylperylene-3,4,9,10-dicarboximide (MePTCDI) is weak due to the indirect character of the corresponding transition
\cite{Hoffmann2000}.

\par

Previous experimental investigations of the archetype material pentacene have provided insight into the exciton dispersion along some crystal
directions, but a complete picture could not be measured due to limitations in the energy resolution \cite{Schuster2007,Knupfer2006}. In this
contribution, we present a comprehensive analysis of the energetically lowest lying electronic excitations of pentacene single crystals using
electron energy-loss spectroscopy (EELS) in transmission with unprecedented energy resolution and at low temperature (20 K). EELS is able to
determine the energy of electronic excitations at finite momentum transfers throughout the entire Brillouin zone of a solid, which can provide
valuable insight into the electronic properties of topical materials
\cite{Schuster2007,Knupfer2000,Marinopoulos2002,Cudazzo2011,Wezel2011,Roth2010}. Our studies allow the exact analysis of the exciton dispersion
in the $\bf{a^*},\bf{b^*}$ reciprocal lattice plane of pentacene, where parallel to the $\bf{b^*}$ direction previous experiments could not
probe the singlet exciton at all. Moreover, the present investigation provides insight into the complex behavior of the higher energy satellite
features, which are discussed in terms of the contribution of vibrations as well as charge transfer excitons.

%%%%%%%%%%%%%%%%%%%%%%%%%%%%%%%%%%%%%%%%%%%%%%%%%%%%%%%%%%%%%%%%%%%%%%%%%%%%%%%%%%%%%%%%%%%%%%%%%%%%%%%%%%%%%%%%%%%%%%

\section{Experimental and sample characterization}

Pentacene single crystals with a high quality and typical crystal dimensions of $5-15$ mm length, $2-5$ mm width, and $0.05-0.25$ mm thickness
were obtained via a directional sublimation of two or three times purified pentacene (Fluka). The crystal growth was carried out at temperatures
between 280$^\circ$ C and 220$^\circ$ C in closed, evacuated pyrex ampoules and a horizontal two-zone furnace was used. The growth procedure
lasted about four to six weeks.

\par

Our EELS investigations require thin, single crystalline pentacene films (thickness $\sim$ 100 nm) which were cut from the largest, flat surface
of a single crystal platelet with an ultramicrotome using a diamond knife. These cut films were attached to standard electron microscopy
grids\cite{Fink1994} and then transferred into the EELS spectrometer.\cite{Fink1989} All EELS measurements were carried out using a 170 keV
spectrometer thoroughly discussed in a previous publication. \cite{Fink1989} At this high primary beam energy only singlet excitations are
possible.\cite{Fink1989} The energy and momentum resolution were 85 meV and 0.03 \AA$^{-1}$, respectively. The EELS signal, which is
proportional to the loss function $Im(-1/\epsilon({\bf q},\omega$)), was determined for various momentum transfers, {\bf q}, parallel to the
directions of the corresponding reciprocal lattice vectors. Moreover, a He flow cryostat allows to choose the sample temperature in the range of
20 K to 400 K.

\par

Prior to the investigation of the electronic excitations, our films were thoroughly characterized in-situ using electron diffraction. These
investigations clearly document that the films are single crystalline as can be seen from Fig. 1 where we depict the electron diffraction
profiles parallel to the reciprocal lattice directions $\bf{a^*}$ and $\bf{b^*}$. Only the respective Bragg diffraction features can be
observed. In addition, we have measured these electron diffraction profiles at temperatures of 300 K and 20 K and determined the length of the
$\bf{a^*}$ and $\bf{b^*}$ reciprocal unit vectors. A comparison to literature data as depicted in Fig. 1 (right panel) reveals that the crystal
structure of our films corresponds to the so-called single crystal polymorph \cite{Mattheus2003a,Mattheus2003b} (polymorph 1 following the
notation of Ref. \onlinecite{Mattheus2003a}) . The temperature dependence of the $\bf{a^*}$ and $\bf{b^*}$ unit vectors arises from the lattice
contraction and is in good correspondence to x-ray diffraction results from the literature as seen in Fig. 1 (right panel). Note that data for
temperatures below 90 K have not been reported so far. Finally, the diffraction experiments were used to orient the momentum transfer parallel
to selected reciprocal lattice directions in the measurements of the electronic excitation spectrum.

\begin{figure}[h]
 \centering
   \includegraphics[width=.6\textwidth]{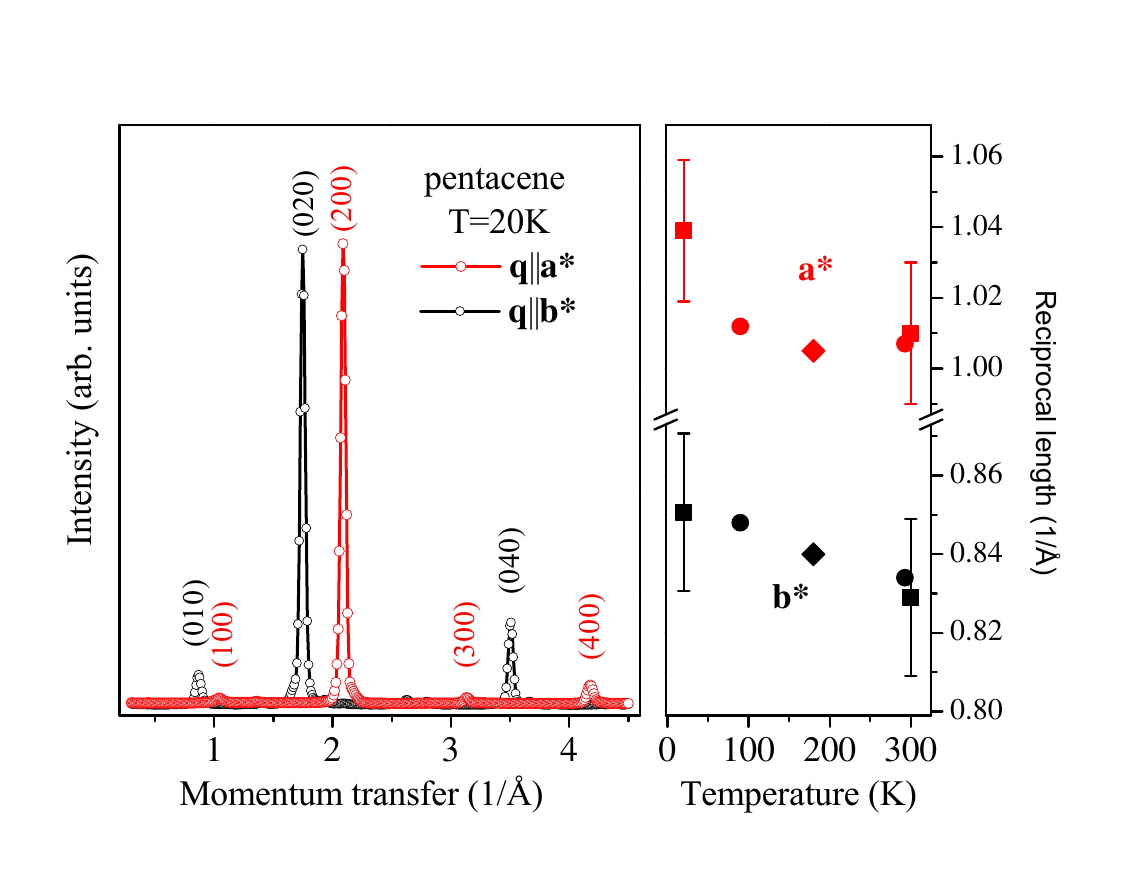}
\caption[]{Representative electron diffraction profile at 20 K (left panel) of the thin pentacene films as studied in this contribution. The
data represent measurements along the two fundamental reciprocal lattice directions and reveal the single crystalline nature of the films. The
right panel shows the reciprocal lattice parameters $a^*$ and $b^*$ from our electron diffraction measurements (square symbols) in comparison to
x-ray diffraction data from the literature. The data for 90 K and 290 K are taken from Ref. \onlinecite{Mattheus2003a}, while those for 180 K
stem from Ref. \onlinecite{Holmes1999}. The very good agreement at 300 K demonstrates that our crystal structure corresponds to the single
crystal polymorph (according to Ref. \onlinecite{Mattheus2003a})} \label{fig1}
 \end{figure}

\par

In order to properly normalize the data taken along different directions in the reciprocal space, we carried out a Kramers-Kronig analysis
(KKA). The raw data first have been corrected by subtracting contributions of multiple scattering processes and elimination of the contribution
of the direct beam  \cite{Fink1989}. The normalization necessary within the KKA analysis has been done using the sum rule for all valence
excitations (for details see Ref. \onlinecite{Fink1989}). The results of the KKA analysis show that data for different directions can be
properly normalized at 2.8 eV to reveal the relative intensities of the corresponding excitations.

\par

Furthermore, as molecular crystals often are damaged by fast electrons, we repeatedly checked our samples for any sign of degradation. In
particular, degradation was followed by watching an increasing amorphous-like background in the electron diffraction spectra and an increase of
spectral weight in the loss function in the energy region below the first excitation feature. It turned out that under our measurement
conditions the spectra remained unchanged for about 12 h. Samples that showed any signature of degradation were not considered further but
replaced by newly prepared thin films. The results from the different films have been shown to be reproducible.

%%%%%%%%%%%%%%%%%%%%%%%%%%%%%%%%%%%%%%%%%%%%%%%%%%%%%%%%%%%%%%%%%%%%%%%%%%%%%%%%%%%%%%%%%%%%%%%%%%%%%%%%%%%%%%%%%%%%%%

\section{Results and discussion}

\begin{figure}[h]
 \centering
\includegraphics[width=.5\textwidth]{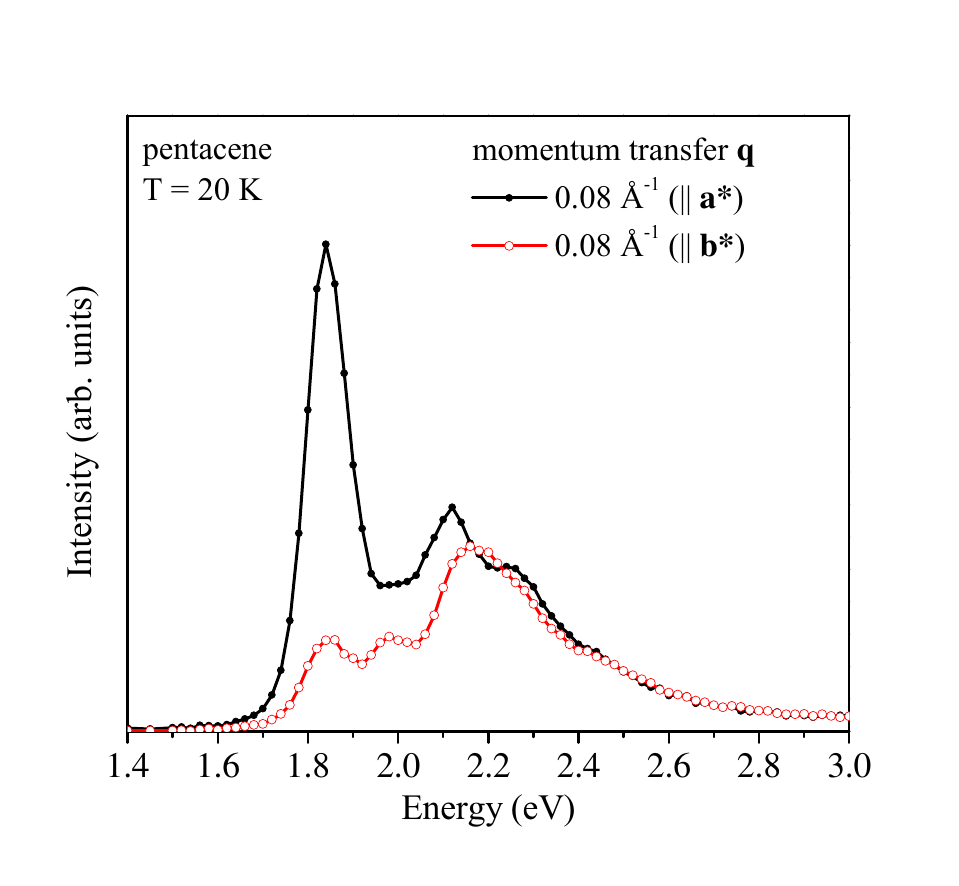}
  \caption{EELS spectra  of single crystalline pentacene films for a small momentum transfer (0.08 \AA$^{-1}$)
  and for momentum vectors $\bf{q}$ parallel to the reciprocal lattice directions $\bf{a^*}$ and $\bf{b^*}$.}
  \label{fig2}
 \end{figure}

We start the presentation of our results with a comparison of the loss function for momentum vectors $\bf{q}$ parallel to the reciprocal lattice
directions $\bf{a^*}$ and $\bf{b^*}$ and a small momentum transfer of 0.08 \AA$^{-1}$. At such a small momentum transfer, which is much smaller
than the size of the Brioullin zone in the $\bf{a^*},\bf{b^*}$ plane of $\ge$ 0.85 \AA$^{-1}$, we probe essentially vertical transitions, the so
called optical limit, i.e. a comparison to optical data is possible. Following the results of our KKA analysis, the data are normalized at 2.8
eV. Fig. 2 presents the highly anisotropic electronic response of pentacene in the two directions studied at a sample temperature of 20 K. For
oligoacenes, it is well known that the lowest singlet exciton is split into two Davydov components, which is a consequence of two non-equivalent
molecules in the triclinic unit cell. For pentacene, these two Davydov components are spilt by about 0.14 eV, and they are predominantly
polarized along the two crystal directions $\bf{a}$ and $\bf{b}$, respectively
\cite{Pope,Zanker1969,prikhotko1969,pirya1989,Faltermeier2006,Dressel2008}. In general, this is also seen in our data as revealed in Fig. 2.
However, it is important to realize the our data are taken along the \emph{reciprocal} lattice directions $\bf{a^*}$ and $\bf{b^*}$, which due
to the triclinic crystal symmetry of pentacene are not parallel to the crystal axes $\bf{a}$ and $\bf{b}$. In particular, the angle between the
crystal axis $\bf{b}$ and its reciprocal counterpart $\bf{b^*}$ amounts to about 14$^\circ$ in pentacene single crystals \cite{Mattheus2003a}.
Therefore, the spectrum in Fig. 2 representing the electronic excitations spectrum along the $\bf{b^*}$ direction reveals two maxima at 1.84 eV
and 1.98 eV, which arise from the two Davydov components, while for a momentum transfer parallel to $\bf{a^*}$ we only observe the lower Davydov
feature. Both curves in Fig. 2 also show higher energy satellites starting above 2 eV. The spectral shape and the energy of these satellite
features is in very good agreement to the results of optical measurements \cite{Faltermeier2006,Dressel2008}. The origin of those will be
discussed in more detail below. In comparison the previously published results from EELS studies of pentacene single crystals
\cite{Schuster2007,Knupfer2006}, the data in Fig. 2 allow the identification of the excitonic features for all momentum vectors in the
reciprocal $\bf{a^*},\bf{b^*}$ plane in detail.

\par

\begin{figure}[h]
 \centering
\includegraphics[width=.34\textwidth]{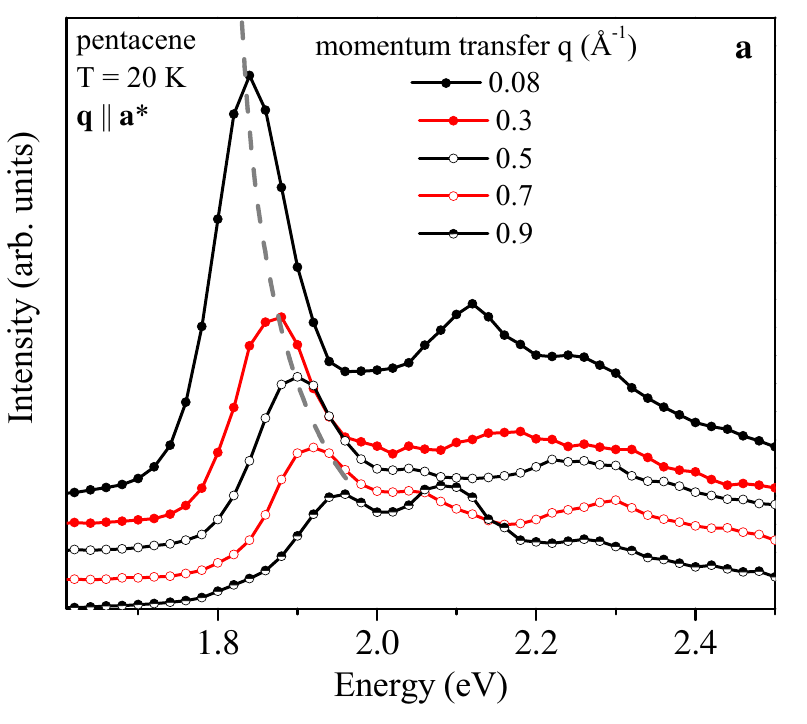}
\includegraphics[width=.34\textwidth]{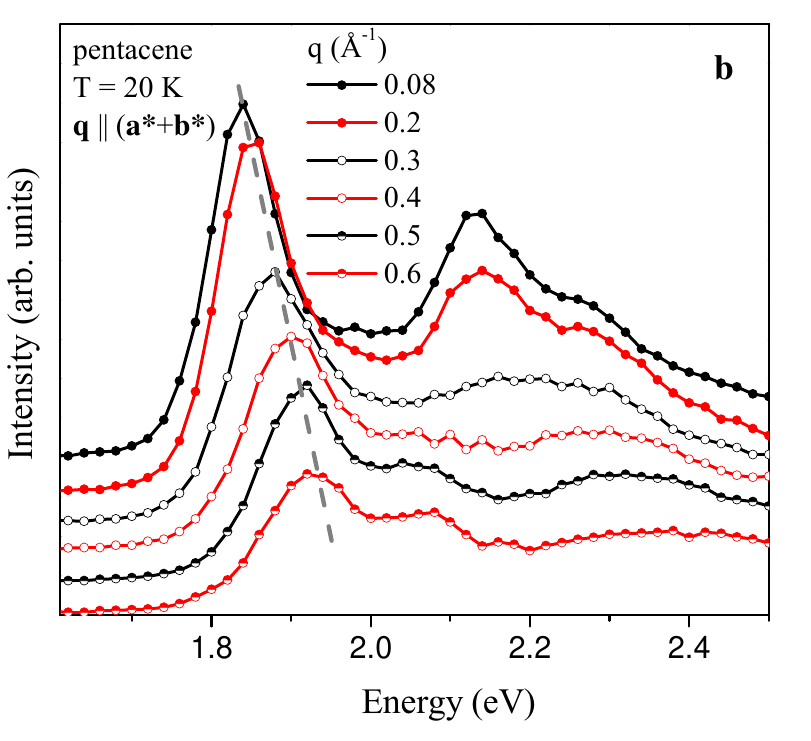}
\includegraphics[width=.34\textwidth]{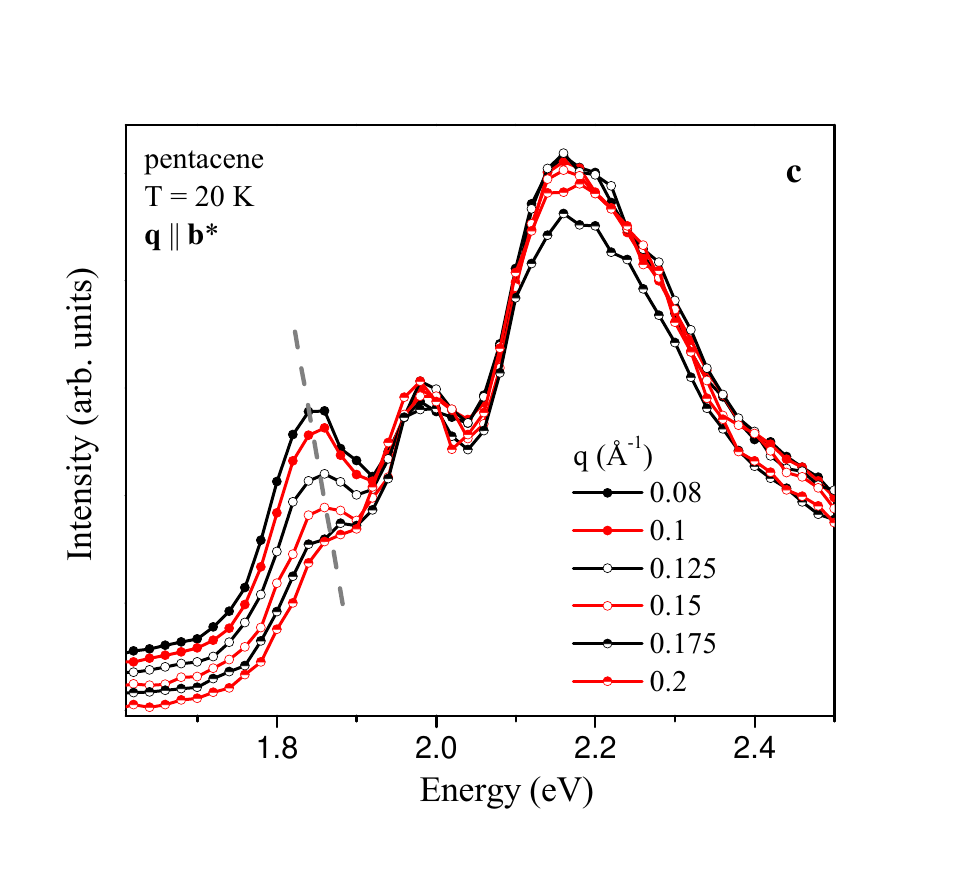}
\includegraphics[width=.34\textwidth]{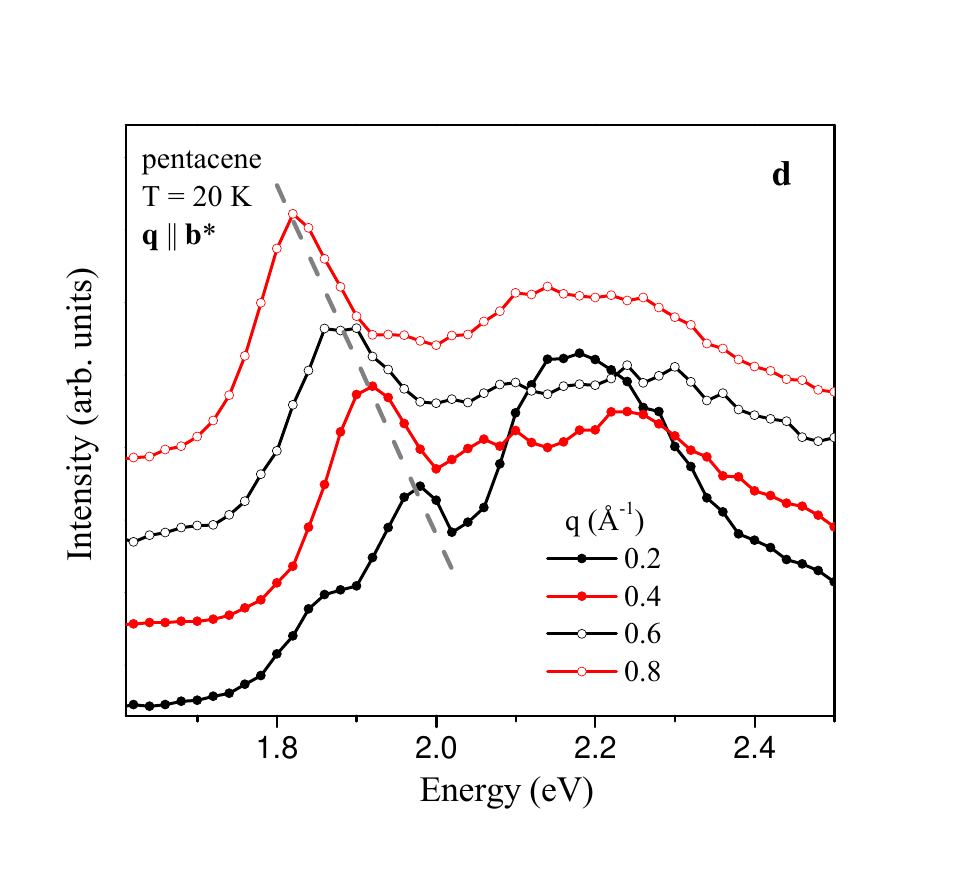}
  \caption{Loss function of single crystalline pentacene as a function of momentum transfer parallel to the reciprocal lattice directions
$\bf{a^*}$ (panel a), $\bf{a^*}+\bf{b^*}$ (panel b), and $\bf{b^*}$ (panel c and d). The size of the momentum transfer is given in \AA$^{-1}$.}
  \label{fig3}
 \end{figure}

The momentum dependence of the excitation spectra for three different directions in the reciprocal $\bf{a^*},\bf{b^*}$ plane are shown in Fig.
3. These data clearly demonstrate the strong momentum dependence of the excitons, i.e. our data provide the exciton band structure. For a
momentum transfer parallel to $\bf{a^*}$ (Fig. 3a), we observe a clear upshift of the exciton with increasing momentum transfer. This upshift is
accompanied by a decreasing spectral weight. Very similar data are observed for a momentum transfer parallel to the diagonal direction
($\bf{a^*}+\bf{b^*}$) (Fig. 3b). The data for $\bf{q}$ $\|$ $\bf{b^*}$ are more complex due to the presence of both Davydov split exciton
components. We therefore present the dispersion data along  $\bf{b^*}$ in two parts. At lower momentum transfers (up to 0.2 \AA$^{-1}$, Fig.
3c), the lowest energy component disperses to larger energies while the second exciton does not show a significant change. Going to larger
momenta the upper Davydov split component disperses negatively reaching the initial energy of the lower component ($q = 0.08$\AA$^{-1}$) at
about 0.8 \AA$^{-1}$ (Fig. 3d).

\par

\begin{figure}[h]
 \centering
\includegraphics[width=.37\textwidth]{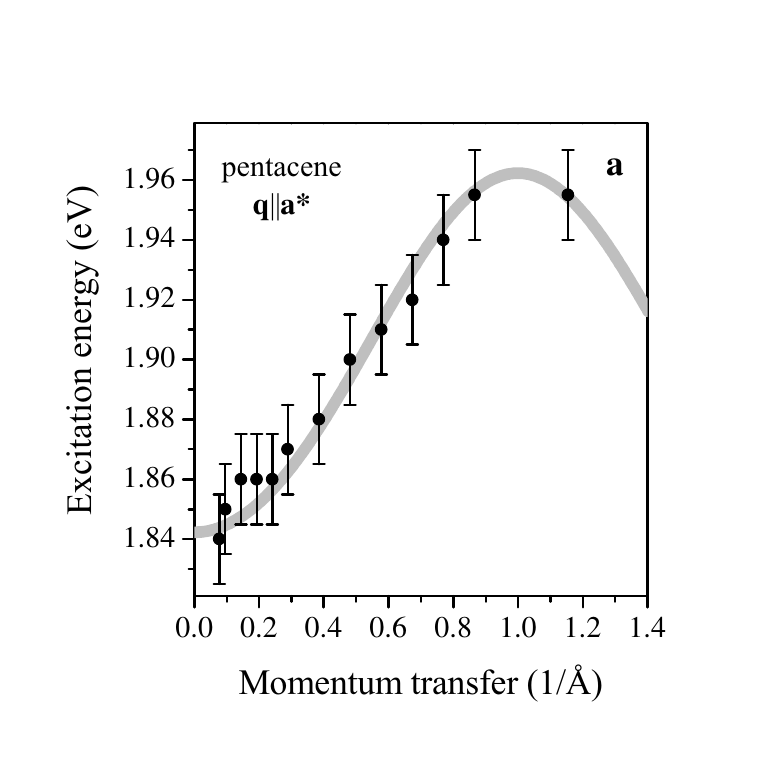}
\includegraphics[width=.37\textwidth]{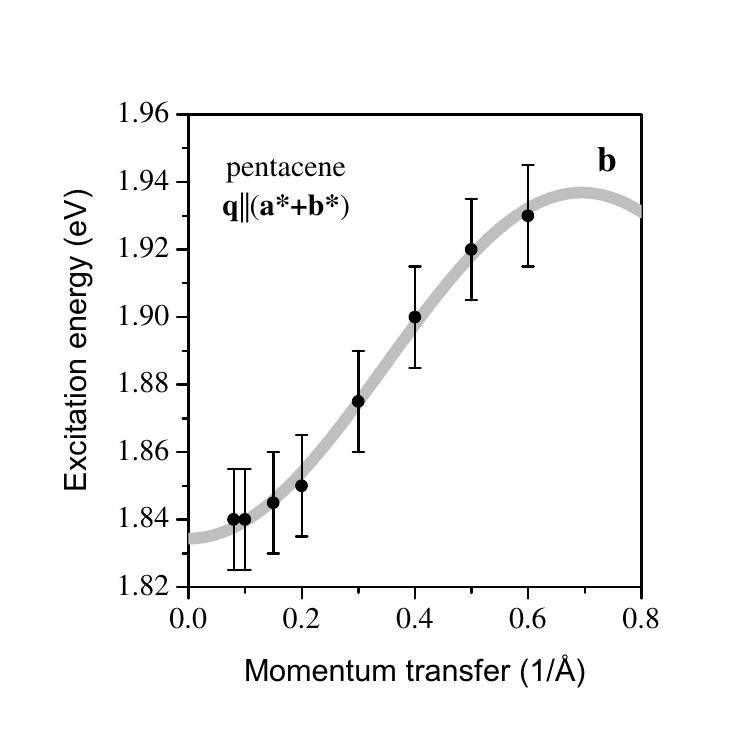}
\includegraphics[width=.37\textwidth]{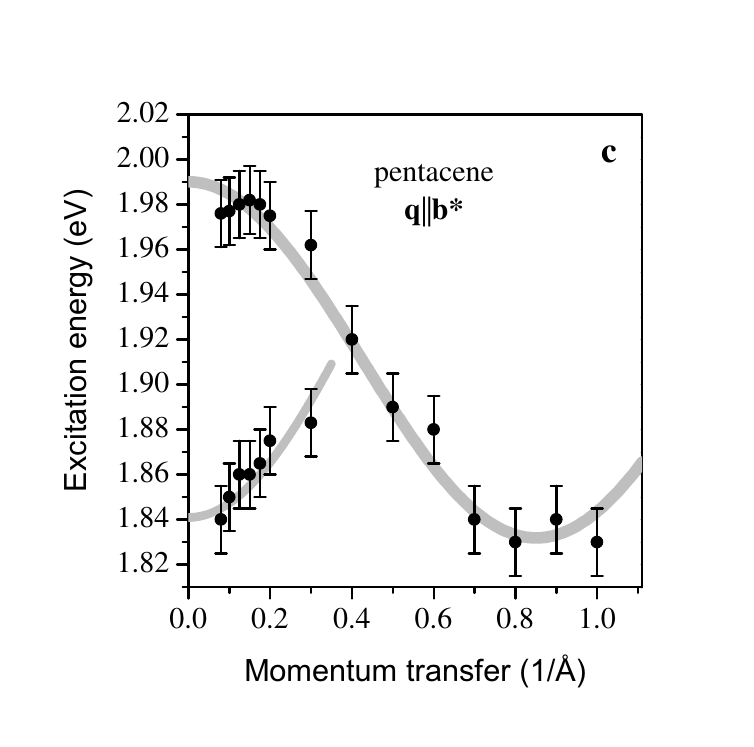}
  \caption{Dispersion of the singlet excitons in pentacene  for momentum transfers parallel to the $\bf{a^*}$ (panel a), $\bf{a^*}+\bf{b^*}$ (panel b),
  and $\bf{b^*}$ (panel c) reciprocal lattice directions. The gray line is intended as a guide to the eye.}
  \label{fig4}
 \end{figure}

The exciton dispersion in pentacene is summarized in Fig. 4 for the three directions discussed above. In general, our data outline that excitons
in single crystalline pentacene can move coherently through the crystal giving rise to a well defined dispersion. The dispersion along
$\bf{a^*}$ and the diagonal ($\bf{a^*}+\bf{b^*}$) is in perfect agreement to the previous results \cite{Schuster2007,Knupfer2006}. The bandwidth
of the exciton bands is about 110 meV. For momentum transfers parallel to $\bf{b^*}$, the lower Davydov component behaves analogous to the other
directions whereas the upper component seems to have a mirror-like dispersion relation. The latter is expected within a picture that relies on
local excitons and their predominant nearest neighbor interaction, since the hopping matrix elements between the two different molecules in the
unit cell, which thereby are mainly responsible for the exciton movement, are the same as for the Davydov splitting \cite{Pope} but enter the
exciton dispersion with opposite sign. However, we emphasize that previous conclusions, that it is impossible to describe the exciton dispersion
using a tight-binding approach based upon nearest neighbor interactions only \cite{Schuster2007}, are fully corroborated. Within such an
approach, which would essentially describe local (Frenkel-like) excitons, three different transfer integrals ($t_a$, $t_b$, and $t_{ab}$) are
considered\cite{Schuster2007,Knupfer2006} (the latter is taken into account as the pentacene crystal unit cell contains two symmetrically
inequivalent molecules, see above). Already the dispersion for momenta parallel to the $\bf{a^*}$ direction can only be modelled when $t_a$ is
negligibly small (for details see Refs. \cite{Schuster2007,Knupfer2006}). $t_{ab}$ then is fixed to describe this direction. Going further to
the other two directions presented here, it is not possible to model both with the only left parameter $t_{b}$. Finally, since the data as
presented in this publication are taken at 20 K while previous results have been measured at room temperature (300 K), the good agreement of the
two data sets indicates an exciton dispersion in pentacene that is virtually independent of temperature below room temperature.

\par

Advanced theoretical considerations of oligoacenes have revealed that the dispersion of the singlet excitons and the observed Davydov splitting
can only be understood when an admixture of charge transfer excitations to the lowest Frenkel-type excitons is taken into account
\cite{Yamagata2011}. Moreover, these calculations indicate a 48 \% (!) charge transfer character of the energetically lowest singlet exciton in
pentacene. As a consequence, it is not surprising that a simple tight-binding approach is insufficient to model the exciton dispersion relations
as it starts from a local, molecular (Frenkel-like) wave function and its nearest neighbor interactions, which is in contrast to the much larger
charge transfer exciton wave function.

\par

Finally, our high resolution data additionally reveal that the spectral width of the excitons is momentum independent throughout the whole
measured momentum range. This is in good agreement with estimates of the exciton binding energies in pentacene of the order of 0.5 eV or larger
\cite{Sebastian1981,Lang2004,Amy2005}, i.e. much larger than the exciton band width. In such a case, the processes limiting the exciton lifetime
can be expected to be momentum independent.

\par

In addition to the well defined exciton features in the range below about 2 eV, there are further excitations visible above this energy for all
spectra presented in Figs. 2 and 3. Moreover, the momentum dependent behavior in this energy range is rather complex. In general, these
excitations above about 2 eV have been discussed in the past on the basis of vibrational satellites and charge transfer excitons. The optical
absorption spectrum of individual pentacene molecules in solution \cite{Zanker1969,Maliakal2004} is characterized by a progression of satellite
features which most likely are caused by the coupling of the excited electronic state with molecular vibrations. This is similar to other
oligoacenes and related molecules. Thus, it is reasonable to expect that at least part of the spectral intensity in our spectra above about 2 eV
is caused by vibrational satellites. We argue however, that on the basis of vibrational effects the intensity distribution and the observed
changes upon momentum variation cannot be understood using vibrational degrees of freedom only. The behavior as seen in our data would imply an
extremely strong momentum dependence of the electron-vibration coupling, which seems rather unlikely in view of the local (intra-molecular)
nature of the relevant vibrations. The impact of the strong mixture of Frenkel and charge transfer degrees of freedom in pentacene however also
changes the higher energy part of our spectra substantially. A 48 \% contribution of charge transfer excitons to the lowest singlet exciton
feature - as predicted \cite{Yamagata2011} - in turn also means that the anti-bonding part of this mixed Frenkel-charge transfer wave function
must have a large Frenkel character, as well. Moreover, previous experiments using electro-absorption indicate that  (charge transfer) excitons
are located at about 0.3 - 0.5 eV higher than the lowest singlet excitation \cite{Sebastian1981}. Consequently, the higher-lying, anti-bonding
component of the mixed wave function also receives from the larger spectral weight of the Frenkel-type excitations, and should show up in the
corresponding spectra. We therefore attribute the spectral structures above about 2 eV as seen in our data to a superposition of vibrational
satellites and excitons of mixed Frenkel-CT character, whereas the latter can have a significant momentum dependence, in analogy to the lower
component as visualized in Fig. 3. Furthermore, also the high energy part of this mixed wave function will couple to the molecular vibrations,
which renders the spectral shape even more complicated. Since the admixture of Frenkel and CT excitons in oligoacenes is expected to depend on
the length of the molecules \cite{Yamagata2011}, a systematic study of the electronic excitation spectra of different oligoacenes might give
further insight into the microscopic origin of the various excitations and their interplay.

%%%%%%%%%%%%%%%%%%%%%%%%%%%%%%%%%%%%%%%%%%%%%%%%%%%%%%%%%%%%%%%%%%%%%%%%%%%%%%%%%%%%%%%%%%%%%%%%%%%%%%%%%%%%%%%%%%%%
\section{Summary}

To summarize, we have studied the low energy singlet excitation spectrum of singly crystalline pentacene films at 20 K as a function of momentum
transfer using electron energy-loss spectroscopy. Our data confirm the substantial anisotropy of the electronic excitation in the
$\bf{a^*},\bf{b^*}$ reciprocal lattice plane. Furthermore, we have determined the full band structure within this plane for the lowest singlet
excitons. These excitons clearly disperse with a total band with of about 110 meV. A comparison to previous measurements taken at room
temperature indicates that the exciton dispersion is rather temperature independent. Following recent theoretical work, the exciton dispersion
can only be understood with the inclusion of charge transfer excitons that significantly mix with the molecular-type Frenkel excitons. The
spectral width of the excitons is momentum independent throughout the entire Brioullin zone in the $\bf{a^*},\bf{b^*}$ reciprocal lattice plane,
i.e. their life time does hardly vary with momentum. Inspection of the excitation features somewhat above the exciton bands indicates that they
cannot be rationalized in terms of vibrational satellites only. Instead, there is evidence that also admixed Frenkel-CT excitons have to be
considered.

\begin{acknowledgments}
We are grateful to R. H\"ubel, S. Leger and R. Sch\"onfelder for technical assistance. Financial support by the Deutsche Forschungsgemeinschaft
within the Forschergruppe FOR 1154 (Project KN393/14) is gratefully acknowledged.
\end{acknowledgments}

%\bibliography{Spiro}


%

\end{document}